\documentstyle[amssymb,preprint,aps,graphicx]{revtex}
\begin{document}

\title{
       High frequency acoustic modes in vitreous Beryllium Fluoride \\
       probed by inelastic X-ray scattering.
      }
\author{
        T.~Scopigno$^{1}$, S.~N.~Yannopoulos$^{2}$, D.~Th.~Kastrissios$^{2}$,
        \\ G.~Monaco$^{3}$, E.~Pontecorvo$^{1}$, G.~Ruocco$^{1}$,
        F.~Sette$^{3}$
        }
\address{
    $^{1}$
    Dipartimento di Fisica and INFM, Universit\'a di Roma
    ``La Sapienza'', I-00185, Roma, Italy.\\
    $^{2}$
    Foundation for Research and Technology - Hellas, Institute
    of Chemical Engineering and High-Temperature Chemical Processes,
    P.O. Box 1414, 265 00 Patras, Greece.\\
    $^{3}$
    European Synchrotron Radiation Facility, BP 220 F-38043,
    Grenoble Cedex, France.
    }
\date{\today}
\maketitle

\begin{abstract}
Inelastic X-ray Scattering measurements of the dynamics structure
factor have bene performed on vitreous Beryllium Fluoride ({\it
v-}BeF$_2$) at $T$=297 K in the momentum transfer, $Q$, range
$Q$=1.5$\div$10 nm$^{-1}$. We find evidence of well defined high
frequency acoustic modes. The energy position and linewidth of the
excitations disperse with $Q$ as $\propto Q$ and $\propto Q^2$,
respectively up to about one half of the first maximum of the
static structure factor. Their magnitude compare favorably with
low-frequency sound velocity and absorption data. The results
indicate worth mentioning similarities of the high frequency
collective dynamics of different network forming glasses such as
{\it v-}B$_2$O$_3$ and {\it v-}SiO$_2$.
\end{abstract}



\section{INTRODUCTION}

The interest in the provocative behavior of the glass-forming
liquid BeF$_2$ has been recently renewed in an effort to relate a
viscosity anomaly to some water-like features, such as the
existence of a negative thermal expansion coefficient region
\cite{hemmati}. Since the first appearance of the $T_g$-scaled
temperature dependence of the viscosity plot for glass-forming
systems, it became clear that liquid BeF$_2$ has an intriguing
property: it shows a crossover between the two extreme strong and
fragile behaviors. In particular, using existing viscosity data
\cite{nev-moy} Angell and co-workers have shown that -in a
$T_g$-scaled plot- the high-temperature limit of BeF$_2$ viscosity
obtained by extrapolation should reach an un-physically small
value \cite{hemmati}. Molecular dynamics simulations have recently
been performed \cite{hemmati} to explain the above
observation in terms of a weak thermodynamic anomaly. In
particular, a density maximum has been predicted to occur at
$T$$\approx$ 2000 K and a density minimum at $T$$\approx$ 1250 K,
a temperature at which a considerable ($\approx$30$\%$) rise in
heat capacity was also predicted.

Despite these intriguing thermodynamics anomalies, BeF$_2$ is much
less studied compared to other network forming glasses like the
oxides SiO$_2$, GeO$_2$, B$_2$O$_3$ etc., owing to
the subtleties that it presents in its purification procedure
(hygroscopicity, toxicity, corrosiveness etc.).

From the experimental point of view the structure of BeF$_2$ has
been elucidated by x-ray \cite{x} and neutron \cite{neu}
diffraction studies supporting the resemblance of BeF$_2$
structure with that of SiO$_2$, i.e. a 3D network of corner
sharing BeF$_4$ tetrahedral units. The room temperature Raman
spectrum of BeF$_2$ has been measured by Galeener et al.
\cite{gal} where a comparison between this glass and other oxides
is attempted. Finally, dynamic properties of BeF$_2$ at low
temperature and ultrasonic frequencies have been carried out
\cite{kratt} showing that {\it i)} between $T$=10 and 200 K,
BeF$_2$ has an acoustic absorption comparable to that of SiO$_2$,
{\it ii)} the temperature dependence of the sound speed of
BeF$_2$ (similar to that of SiO$_2$) shows anomalous behavior. In
particular, in contrast to the monotonically decreasing sound
speed with temperature rise which is usually found in glasses
(B$_2$O$_3$, GeO$_2$, Zn(PO$_3$)$_2$), the sound
speed of BeF$_2$ (and of SiO$_2$ as well) exhibits an initial
drop followed by an upturn above ~50 K.

In this paper, we present the first experimental determination of
the acoustic properties of BeF$_2$ in the {\it THz} frequency
range, by means of Inelastic X-ray Scattering (IXS). In
particular, we have measured the dynamic structure factor,
$S(Q,\omega)$, at room temperature ($T_g$=598 K) {\it i)} as a
function of energy $E$ for fixed values of $Q$ in the range $1
\div 8$ nm$^{-1}$; {\it ii)} as a function of $Q$ for the fixed
energies of 0 and 7 meV. The obtained data allow us to extract
some information about the high frequency sound dispersion and
attenuation properties for the longitudinal acoustic excitations of
this network forming glass. Specifically, we find that {\it i)}
Well defined high frequency acoustic modes exist and propagate in
the glass; {\it ii)} The excitation frequency $\Omega(Q)$
disperses linearly with $Q$ ($\Omega(Q)=vQ$); {\it iii)} The
excitation width $\Gamma(Q)$ (FWHM) increases quadratically with
$Q$, ($\Gamma(Q)=DQ^2$); {\it iv)} The sound speed, $v$, and
the sound energy absorption coefficient, $\alpha =2\pi \Gamma /v$,
compare favorably with literature low-frequency ultrasonic data \cite{note1}.
Finally, a comparison of the sound absorption data of BeF$_2$ with
that of SiO$_2$ indicates striking similarities in the
behaviour of the high frequency collective dynamics of these
two systems.

\section{EXPERIMENTAL DETAILS}

Beryllium fluoride is a substance that presents considerable difficulties in
its purification procedure. This stems from the fact that it is extremely
hygroscopic, toxic and corrosive for conventional containers like fused silica
tubes. For the above reasons the whole material handling operation took place
in an inert atmosphere (nitrogen-filled) glove box with a water content less
than 2 ppm while the material was melted in gold-plated silica tubes of 8 mm
inner diameter. The BeF$_2$ starting material was purchased from Alfa Aesar
Chemical Co. with nominal purity 99.5 $\%$ which was not enough to obtaine
a transparent glass/melt free of black spots. Thus, BeF$_2$ was first sublimed
under dynamic high vacuum and high temperature in graphite tubes. The material obtained
after sublimation contained pure and contaminated parts from the container. The
proper amount of pure BeF$_2$ was selected and placed into the gold-plated
cylindrical silica crucibles with flat bottom. The crucibles with the pure
BeF$_2$ were placed in a quartz cell that was evacuated,
partially filled with Ar, and flame sealed. This cell was then transferred into a
furnace and heated up to the softening point of the glass (1100 K) for 1-2
hours to obtain complete homogenization. The melt was cooled down to room
temperature in a controlled way in order to obtain a transparent glass, free
of internal stresses. With this procedure we were able to obtain cylindrical
BeF$_2$ glass samples with a length of about 1 cm, comparable with the
absorption length for 21.7 keV x-rays. Characteristic temperatures of
BeF$_2$ are: glass transition temperature $T_g=$ 598 K and melting point of
both the quartz-like and cristobalite-like crystalline modifications $T_m=828$
K.

Data for the BeF$_2$ glass have been collected at room temperature ($T$=297 K) at
the IXS beamline ID28 of ESRF \cite{noiV,noiM}. The experiment has been
performed at fixed exchanged wavevector over a $Q-$ region $1.5 \div 10$
nm$^{-1}$ with a resolution (FWHM) of $\delta Q \approx 0.35$ nm$^{-1}$. The
overall energy resolution (FWHM) has been set to $\delta E=1.5$ meV utilizing
the (11 11 11) reflection for the Si monochromator and crystal analyzers.
A five-analyzers bench, operating in horizontal scattering geometry, allowed
us to collect simultaneously spectra at five different values of $Q$. Each
energy scan ($-35 < E < 35$ meV) took about 300 minutes, and several scans have

been accumulated for a total integration time of about 500 seconds/point. Measurements
at constant energy have also been conducted, scanning over the scattering angle.
In this case the energy resolution was the same as in the fixed $Q$ scans,
while the $Q$ resolution was increased, $\delta Q \approx 0.17$ nm$^{-1}$, by using
tunable-width slit placed on the scattered beam path.

\section{DATA PRESENTATION AND DISCUSSION}

\begin{figure}[p]
\centering \vspace{-.4cm}
\includegraphics[width=.47\textwidth]{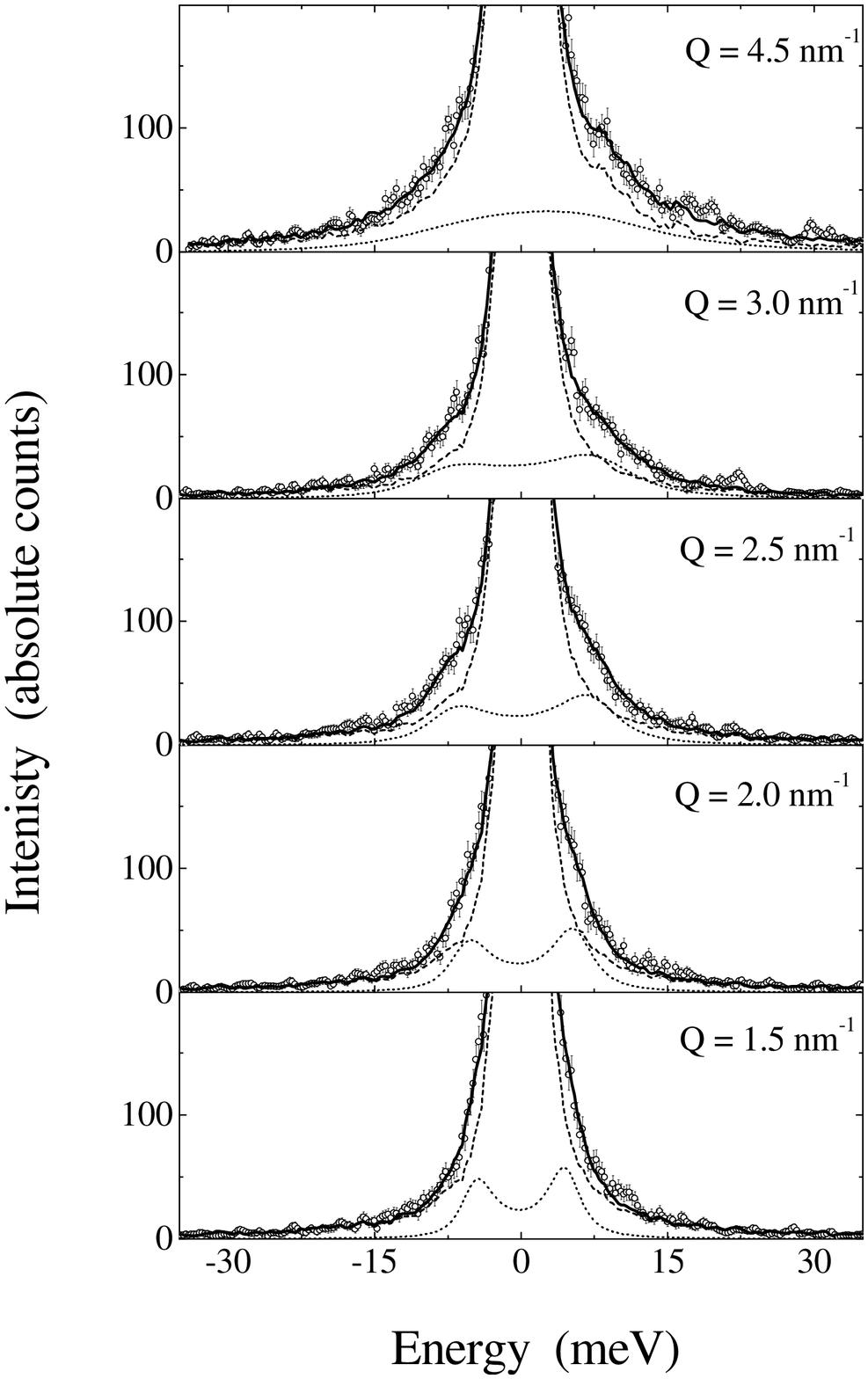}
\caption{Selected IXS energy spectra of the BeF$_2$ glass at room temperature and
fixed values of the exchanged wavevector (open dots $\circ$). The solid lines
represent the best fits with Eq. \ref{fitfunction}, while dashed and dotted
lines account for the elastic and inelastic contributions, respectively, as
determined by the fitting procedure.} \label{panel}
\end{figure}

Fig.~\ref{panel} illustrates a selection of experimental spectra
accumulated at the indicated $Q$ values as a function of the exchanged
energy $E$. Since the incident flux on the sample slightly varies
during the acquisition of each scan, the data have been normalized
to a monitor signal for each frequency value, and then multiplied by
the average monitor. The data of
Fig. 1 bear a close resemblance with those collected from other
network forming glasses, such as SiO$_2$ \cite{bensil,massil} and
B$_2$O$_3$ \cite{matic}. A strong elastic peaks dominates the
spectra, and the inelastic features appear only as weak shoulders
on the tail of the resolution-broadened elastic line.
In order to extract quantitative information on
the excitations giving rise to the inelastic signal, we fitted the
the experimental data with a model function for the $S(Q,E)$ convoluted
with the instrumental resolution.

In fact, the actual experimental intensity, $I(Q,E)$, is proportional to the
convolution of the dynamic structure factor, $S_Q(Q,\omega)$,
($\omega=E/\hbar$), with the instrumental resolution function $R(E)$.

\begin{equation}
I(Q,\omega )=e(Q)\int d\omega ^{\prime }S_Q(Q,\omega ^{\prime
})R(\omega -\omega ^{\prime })  \label{convo}
\end{equation}

\noindent where $e(Q)$ contains the efficiency
of the analyzers, the atomic form factors
and other angular-dependent correction factors. The true, quantum,
$S_Q(Q,\omega)$ can be approximately related to its classic
counterpart by

\begin{equation}
S_Q(Q,\omega) \approx \frac{\hbar \omega / KT }{1-e^{-\hbar \omega / KT }}
S(Q,\omega)
\end{equation}

An useful expression for the (classical) dynamic
structure factor can be obtained by recalling that its Fourier transform
in the frequency domain, i.e. the density fluctuation correlation function
$F(Q,t)=\langle \delta\rho_{\bar Q}(t)\delta\rho^*_{\bar
Q}(0)\rangle$, obeys a generalized Langevin equation
\cite{baluc}:
\begin{equation}
    \label{lang}
    \ddot F(Q,t)+\omega^2_o F(Q,t)+
    \int_o^t m(Q,t-t') \dot F(Q,t') dt' =0
\end{equation}
where $\omega_o$ is a parameter related to the static structure factor
$S(Q)$ \cite{baluc}, and $m(Q,t)$ is the "memory
function". In this
``exact'' expression all the difficulties associated to the
calculation of $F(Q,t)$ have been transferred to the
determination of $m(Q,t)$, with the advantage that the first
two sum rules for $S(Q,\omega)$ are automatically satisfied
\cite{baluc}. By Fourier transforming Eq.~(\ref{lang}), it is
easy to show that:
\begin{eqnarray}
    \frac{S(Q,\omega)}{S(Q)}=\frac{1}{\pi} \frac{
    \omega_o^2 \; m^{\prime }(Q,\omega)}
    {\left[ \omega^2-\omega _o^2+
    \omega m^{\prime \prime }(Q,\omega)\right]^2
    +\left[ \omega m^{\prime }(Q,\omega)\right] ^2}
\label{sqwgenerale}
\end{eqnarray}
where $m^{}(Q,\omega)$=$m^{\prime }(Q,\omega)$+$i
m^{\prime\prime}(Q,\omega)$ is the time Fourier transform of the
memory function $m(Q,t)$. In the $\omega \tau_\alpha
>>$1 limit, a limit that is certainly valid in the present case of a
glassy sample, the memory function can be approximated by the sum
of a constant term, $\Delta^2(Q)$, -which reflects the frozen
$\alpha$-process- plus a function showing a very fast decay at
short times. The latter contribution to the memory function -often
referred to as ``microscopic'' or ``instantaneous''- is usually
represented as a delta-function with area $2\Gamma(Q)$. Therefore:
\begin{equation}
m(Q,t)=2\Gamma(Q)\delta(t)+\Delta_\alpha^2(Q) \label{memDHO}
\end{equation}
and hence Eq.~(\ref{sqwgenerale}) reads as:
\begin{eqnarray}
\frac{S(Q,\omega)}{S(Q)}=\left [ f_Q \delta(\omega) + \frac{1-f_Q}{\pi}
\frac{\Omega^2(Q)\Gamma(Q)} {(\omega^2-\Omega^2(Q))^2+\omega^2\Gamma^2(Q)}
\right ] \label{DHO}
\end{eqnarray}
where $\Omega(Q)$=$\sqrt{\Delta_\alpha^2(Q)+\omega_o^2}$, and
$f_Q$=1-$\omega_o^2/\Omega^2(Q)$ is the non-ergodicity factor.
The expression in Eq.~\ref{DHO} is the sum of an elastic line (the
frozen $\alpha$ process) accounting for a fraction $f_Q$ of the
total intensity, and of an inelastic feature which is formally
identical to a Damped Harmonic Oscillator (DHO) function. The
parameter $\Omega(Q)$ coincides with the maximum of the
longitudinal current correlation function,
$J(Q,\omega)=({\omega^2}/{Q^2}) S(Q,\omega)$, and it is related to
the apparent sound speed value $c_l(Q)=\Omega(Q)/Q$, while $\Gamma
(Q)$, the excitation width, is related through its low $Q$ value
to the acoustic absorption coefficient $\alpha={2*\pi
\Gamma(0)}/{c_l(0)}$.

The model function, $I^M(Q,\omega)$ utilized to fit the
experimental data will be, therefore

\begin{eqnarray}
&I^{M}(Q,\omega)=a(Q)\;\int\frac{\hbar \omega' / KT }{1-e^{-\hbar
\omega' / KT }} R(\omega- \omega') \times \label{fitfunction} \\
&\left[f_Q \delta(\omega')+\frac{1-f_Q}{\pi}\frac{\Omega^2_Q
\Gamma_Q}{(\omega'^2-\Omega^2_Q)^2+\omega'^2 \Gamma^2_Q}\right]
d\omega ' \nonumber
\end{eqnarray}
which, apart from an intensity factor $a(Q)=e(Q)S(Q)$, contains
three free shape parameters: $\Omega_Q$, $\Gamma_Q$ and $f_Q$.

In Fig.~\ref{panel}, together with the raw data (circles), we
report the the elastic contribution (dashed lines), the fits to the
data (full lines) and the inelastic contribution (dotted lines).
This latter contribution clearly shows a dispersing behavior
despite the considerable sound attenuation
($\Gamma(Q^*)=\Omega(Q^*)$ at $Q^*\approx 3$ nm$^{-1}$).

\begin{figure}[p]
\centering \vspace{-.5cm}
\includegraphics[width=.45\textwidth]{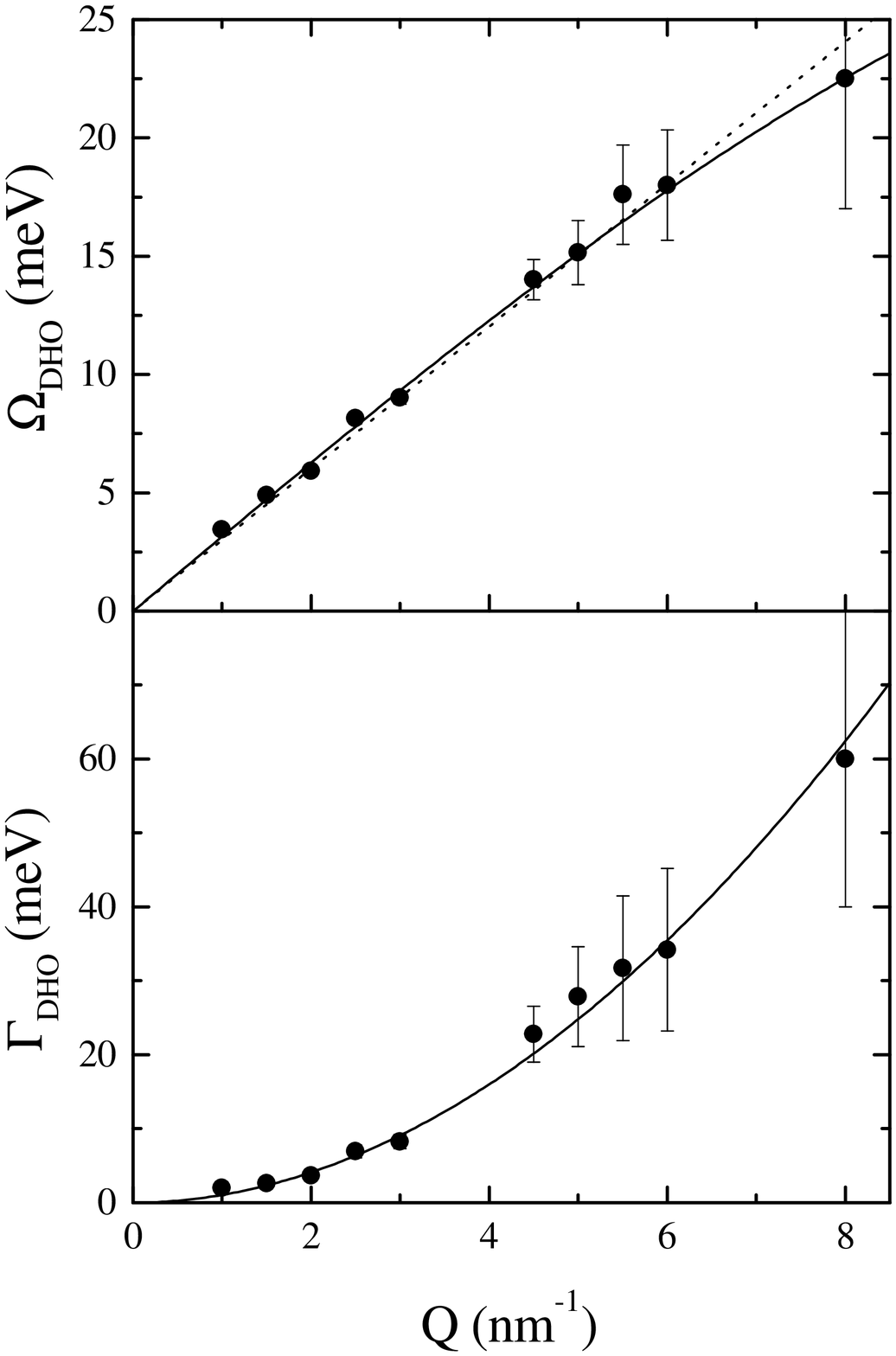}
\caption{$Q-$dependence of the sound frequency and attenuation.
Upper panel: experimental sound dispersion (full dots) and best
fit through a sine function (full line). The adiabatic sound
speed, measured by ultrasounds, is also reported (dotted line).
Lower panel: Acoustic attenuation as determined by the Brillouin
linewidth (full dots). The solid line is a power law fit to the
experimental data and gives $\Gamma_{DHO}=1.05Q^{1.96}$ }
\label{disp}
\end{figure}

In Fig.~\ref{disp} (upper panel) we report the best fit values of
$\Omega(Q) $ (dots) along with the dispersion obtained in the low
frequency (ultrasound) limit (dotted line, corresponding to the
adiabatic sound speed $c_0=4570$ m/s \cite{kraspeed}). To
represent the $Q$-dependence of $\Omega(Q)$ we utilized the
expression $\Omega(Q)=c_l(0)Q_0sin(Q/Q_0)$ in order to keep into
account for the bending of the dispersion curve associated to the
existence of a quasi-Brillouin zone. This procedure yielded for the
sound speed $c_l=4790 \pm 50$, a value that, as in the case of
vitreous silica \cite{massil}, exceeds its low frequency value. The higher
value of the high frequency sound velocity could be the signature
of the crossing of some relaxation process. Such a process can not
be the structural relaxation, since in the glass it lies in the
sub-$Hz$ region, while it can be
ascribed to the action of a microscopic process -produced by the
topological disorder that is present in the glassy phase, as
demonstrated in the case of simulated monatomic glasses
\cite{har,litiosim}. This hypothesis implies that the assumption
of an instantaneous decay of the memory function in eq.
\ref{memDHO} may be inadequate; unfortunately, however, the low magnitude of the
inelastic to elastic ratio does not allow us for a more realistic ansatz
for the memory function shape. The excitation width
$\Gamma_Q$ is reported as a function of $Q$ in the lower panel of
Fig.~\ref{disp}. Its $Q$-dependence is compatible with the law
already found in all the glasses where the dynamic structure
factor has been measured in the mesoscopic region: $\Gamma_Q =D
Q^2$. Actually, a fit to the data with the law $\Gamma_Q =D
Q^\gamma$ (full line) results in $\gamma$=1.96$\pm$0.05 and
$D$=1.05$\pm$0.05 meV/nm$^{-2}$.

\begin{figure}[p]
\centering
\includegraphics[width=.45\textwidth]{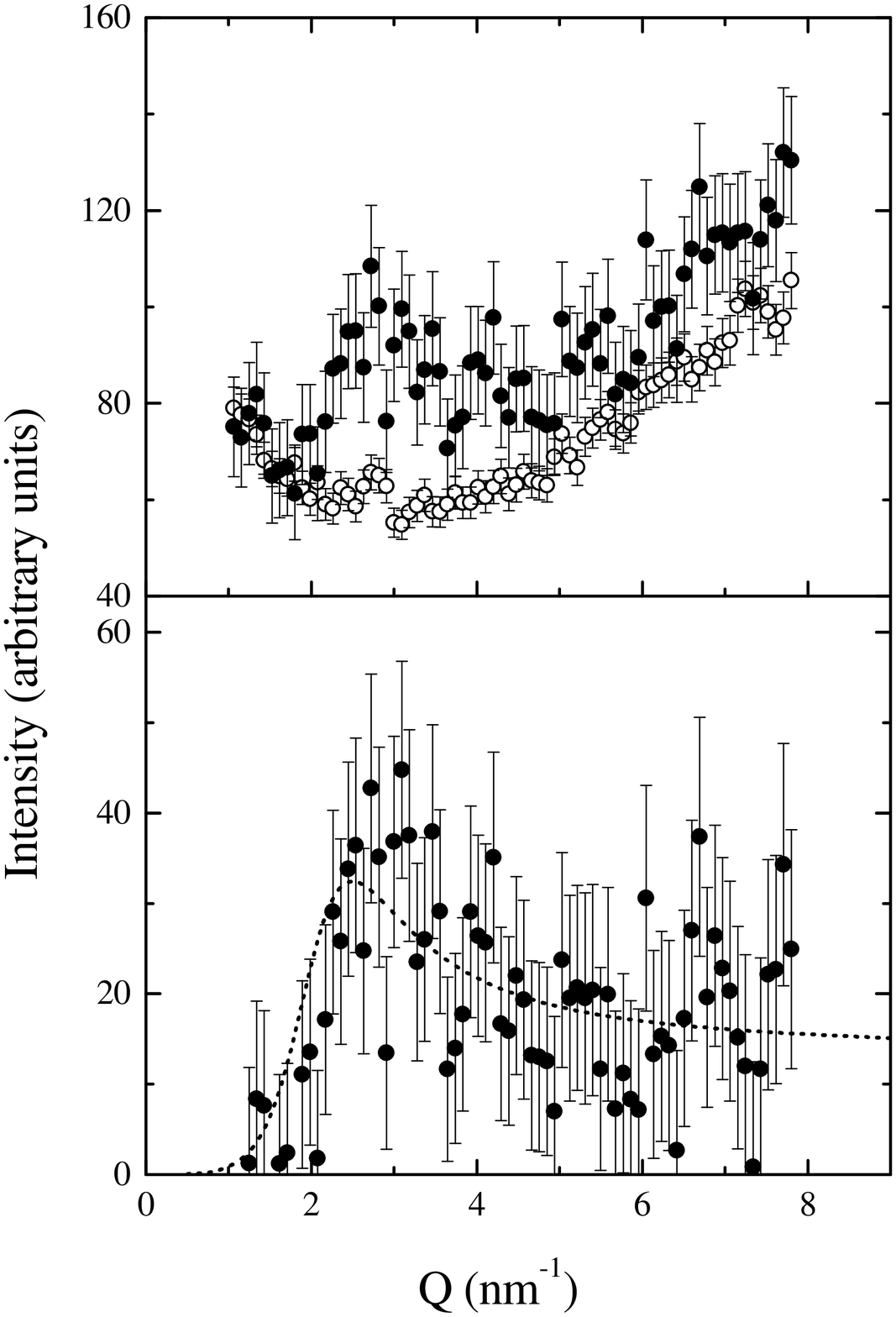}
\caption{Fixed energy IXS $Q-$scan. Upper panel: inelastic ($I(Q,E=7$ meV),
full dots) and elastic ($I(Q,E=0$ meV), open dots). Lower panel: genuine
inelastic signal determined as the difference of the above reported scans
(normalized according to Ref. \protect \cite{pilla}), compared to a theoretical
DHO with parameters determined as discussed in the text. } \label{qscan}
\end{figure}

In order to further illuminate the previous results, the dynamic
structure factor has also been measured at a fixed exchanged
energy value as a function of the momentum transfer. In the
constant $Q$ scans (as those reported in Fig.~1), indeed, the
inelastic signal always appears as a shoulder of the resolution
broadened central line, so that the choice of an appropriate model
for the $S(Q,\omega)$ can be, in principle, neither easy nor
unique \cite{pilla}. In a constant $E$ scan, on the contrary, the
inelastic signal is more clear, since the (resolution
broadened) elastic contribution appears as an almost flat,
$Q$-independent, background. Moreover, the parameters determined
trough the best fit to the energy scan data can be utilized to build-up a curve that can be compared to
the constant $E$ experimental data \cite{pilla}. In
Fig.~\ref{qscan} we present, as an example, the result of such a
comparison for a scan at $E$=7 meV: in the upper panel we show the
raw experimentally measured elastic, $I(Q,E=0 $ meV$)$, and
inelastic, $I(Q,E=7 $ meV$)$ data. Obviously, the latter data
contain an elastic contribution as a consequence of the finite
energy resolution. In order to subtract this elastic contribution
we used a series of spectra at constant $Q$ taken between $Q$=23
and 32 nm$^{-1}$. After aligning and scaling the experimentally
determined resolution function to the elastic peaks in the spectra
at high $Q$, we estimate the relative intensity between the
elastic and inelastic signals at the energy transfers utilized in
the constant-$E$ spectra. The elastic to inelastic intensity
ratio has been obtained at $Q$=13, 16, 19, and 22 nm$^{-1}$.
These ratios allow us to put in the correct relative scale the
spectra taken at $E$=7 meV and at $E$=0. This normalization
procedure is used to derive the inelastic part of the $S(Q,E)$ by
the subtraction of the normalized elastic contribution from the
total scattered intensity. The difference spectrum at $E$7 meV is
reported in the lower panel of Fig.~\ref{qscan} (circles) together with the
error bars as derived from the counting statistics. In this
spectrum the existence of a defined Brillouin peak is highly
emphasized. The dashed line in the lower panel of Fig.~\ref{qscan}represents
the function {\it predicted} on the basis of Eq.~\ref{fitfunction}, with
$\Omega_Q=c_l(0)Q_0sin(Q/Q_0)$ and $\Gamma_Q=D Q^\gamma$ (as
derived from Fig.~\ref{disp}), where the only adjustable parameter
is now an intensity factor since $c_l(0)$, $Q_0$, $D$ and $\gamma$
are determined from the fit results of the energy scans. Although the peak
position turns out to be slightly underestimated, the proposed fitting
model appears to capture the main features of the experimental
data.

The convergent results obtained by analyzing the two independent data sets, i.~e. the constant-$Q$ and -$E$ scans, allow us to establish the
appropriateness of the approximations introduced in the memory function of the
density fluctuations, and, more specifically, of the results reported in
Fig.~\ref{disp} for $\Omega_Q$ and $\Gamma_Q$.

\begin{figure}[h]
\centering
\includegraphics[width=.95\textwidth]{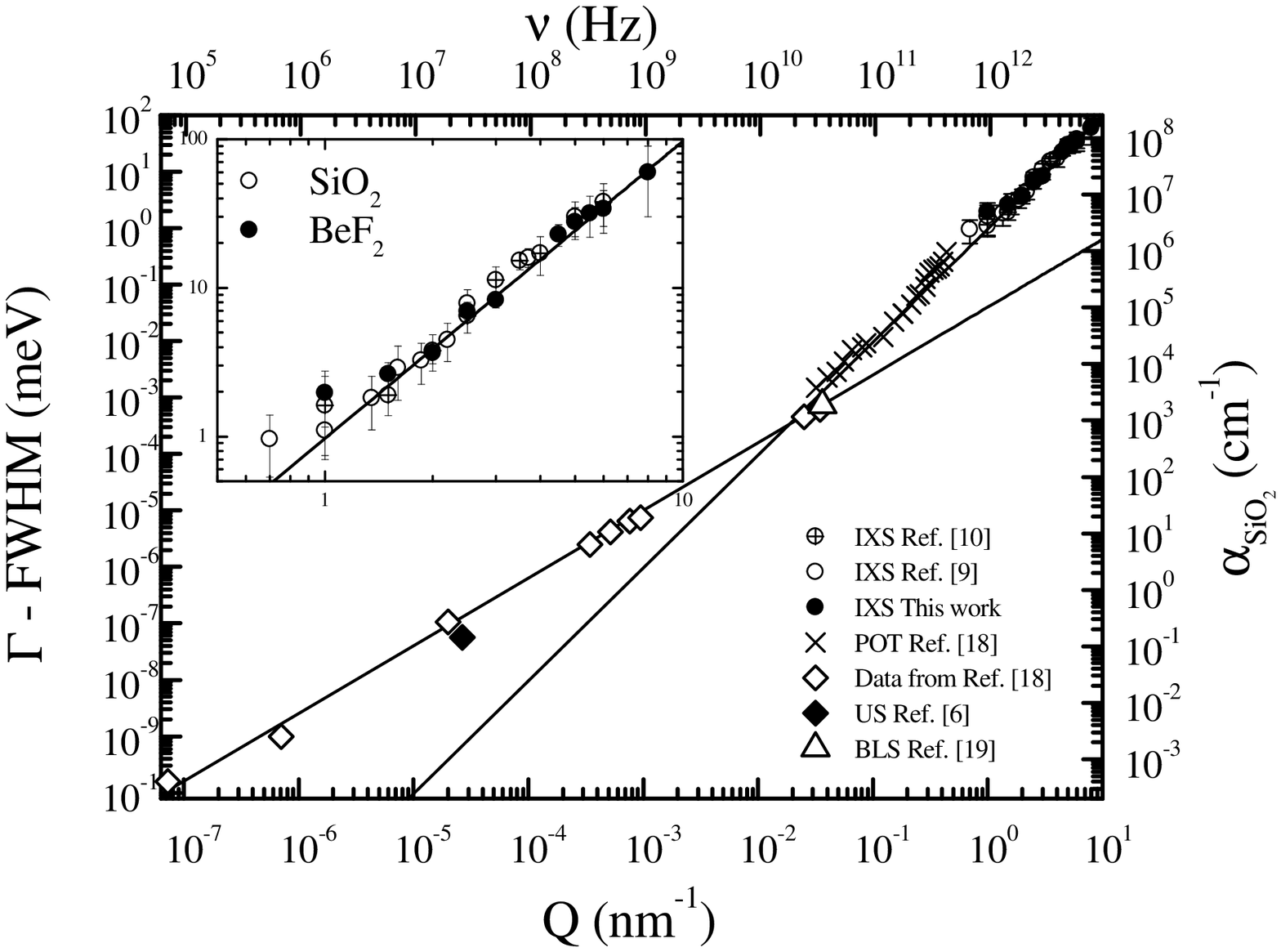}
\vspace{-10.2cm} \caption{Sound attenuation of BeF$_2$ and SiO$_2$ determined
with different techniques (the right scale is for SiO$_2$ data only). The full
lines are $Q^2$ and $Q^{1.3}$ power laws. The inset details the IXS frequency
window} \label{att}
\end{figure}

One of the most intriguing topics, as far as the glassy dynamics
is concerned, is the frequency (or $Q$) dependence of the sound
attenuation. Despite that in the IXS window (i.~e. at $Q$ values
ranging from $\approx$1 to $\approx$10 nm$^{-1}$, or energies
ranging from $\approx$5 to $\approx$50 meV) a $Q^2$ law
describes the excitation broadening, at lower frequencies the
situation becomes less clear due to the presence of relaxation
processes. An interesting comparison between BeF$_2$ and SiO$_2$
is presented in Fig. \ref{att} where we report the $\Gamma(Q)$
parameter for these two glasses over a wide range of frequencies
as derived from IXS and other literature data. Both glasses exhibit a
$Q^2$ behavior above 35 GHz, while below this value a $Q^{1.2}$
power law well represents the available experimental data for SiO$_2$
\cite{zhu}.

\begin{figure}[p]
\centering
\includegraphics[width=.45\textwidth]{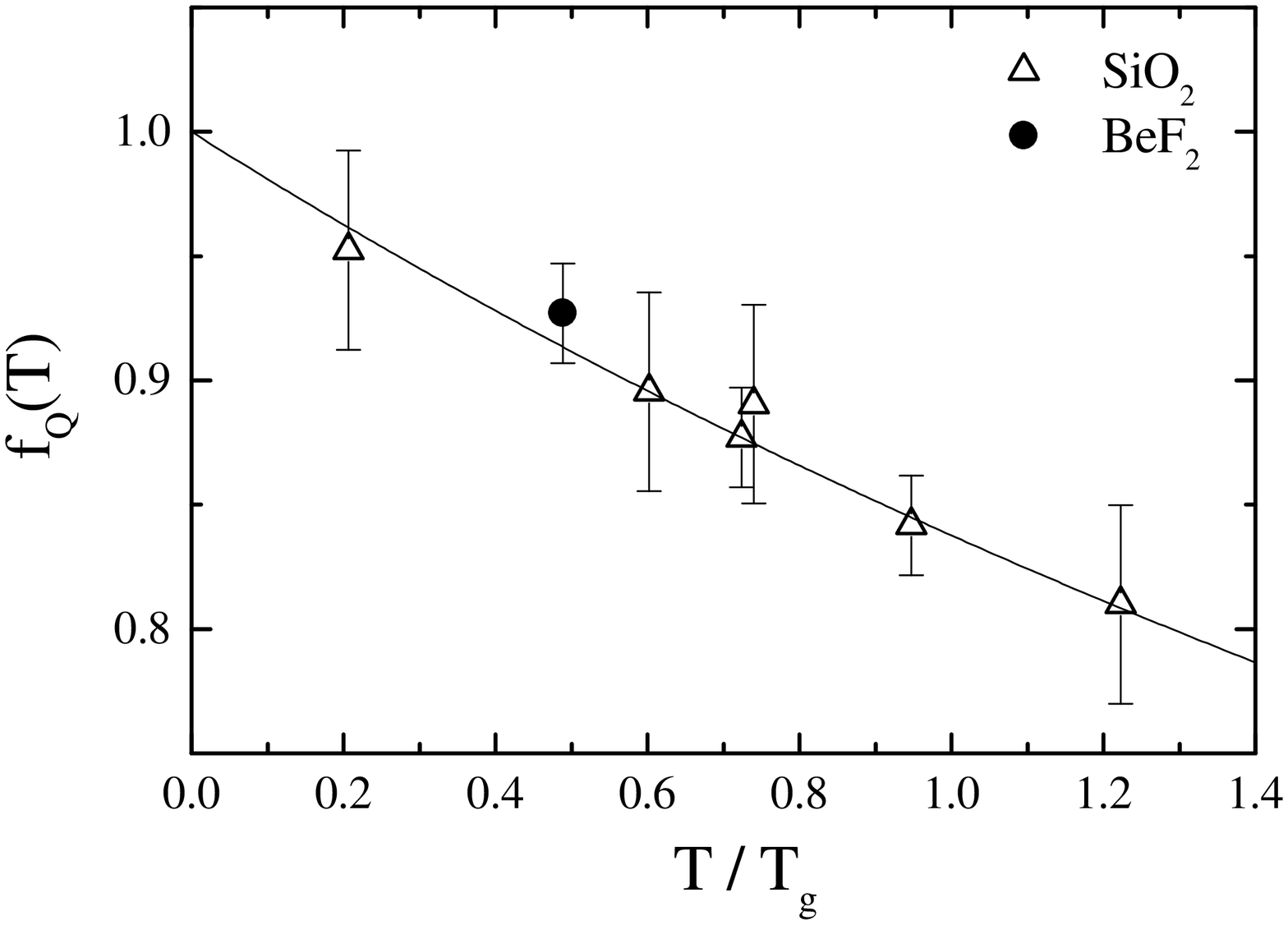}
\vspace{-4.5cm} \caption{Non-ergodicity factor of BeF$_2$ and
SiO$_2$ in a $T_g$ scaled plot. The reported data have been
determined for momentum transfers ranging from $1.6$ to $2.5$
nm$^{-1}$, a region where $f(Q)$ turns out to be rather constant.
The full line is a guideline to the eyes only.} \label{fq}
\end{figure}

Finally, in Fig. \ref{fq}, we present the non ergodicity factor as
determined by IXS in vitrous silica and Beryllium Fluoride.
Indicating with $I_{el}=S(Q)f(Q)$ and $I_{inel}=S(Q)[1-f(Q)]$ the
integrated elastic and inelastic spectral contribution (determined
by the fit), the non ergodicity factor reads as
$f(Q)=\frac{I_{el}}{I_{inel}+I_{el}}$. As shown in the figure, the
$f(Q)$ of BeF$_2$ is in good agreement with the SiO$_2$ data when
reported as a function of $T/T_g$. The results reported in
Figs.~\ref{att} and \ref{fq} imlpy the existence of an
interesting analogy between these two network forming glasses as
far as their high frequency dynamics is concerned.

\section{CONCLUSIONS}

In conclusion, a room temperature IXS study of glassy BeF$_2$ has been undertaken in
the present work. In accordance with all previous studies in
strong and fragile glasses, evidence has been presented for well-defined
propagating (high frequency) acoustic modes, whose frequency
position and linewidth scale as $\propto Q$ and $\propto Q^2$,
respectively. The longitudinal speed of sound for glassy BeF$_2$ has
been estimated to exceed its low frequency (ultrasonic) limit by
almost 5$\%$; a case analogous to that found in studies of
vitreous silica. The extrapolation of the high-frequency linewidth
conforms nicely with the value obtained from ultrasonic studies,
and exhibits a scenario similar to that of vitreous silica.
Another similarity which deserves further study is that the
temperature dependence (in a T$_g$-scaled plot) of the non-ergodicity
factor, as determined from the ratio of the elastic to the total
scattered intensity, follows the behavior exhibited by SiO$_2$
(Fig.\ref{fq}) while for less strong glasses the drop of $f(Q)$ is
much faster with increasing temperature. Unfortunately, for BeF$_2$
it is up to now available only one point in such a $T_g$-scaled plot,
which however coincides with the $f(Q)$ data
for silica. Further temperature-dependence studies on BeF$_2$ are
expected to shed more light on this issue.

\section{ACKNOWLEDGEMENTS}

We kindly acknowledge M. Krisch and the staff of the ID28 beamline at the ESRF
for valuable help and assistance during the measurements, in particular the
local contacts M. Lorenzen and R. Verbeni.

\newpage

\vspace{7.cm}
Fig. 1 - T. Scopigno et al, ``High frequency acoustic modes...''

\newpage

\vspace{7.cm}
Fig. 2 - T. Scopigno et al, ``High frequency acoustic modes...''

\newpage

\vspace{7.cm}
Fig. 3 - T. Scopigno et al, ``High frequency acoustic modes...''

\newpage

\vspace{7.cm}
Fig. 4 - T. Scopigno et al, ``High frequency acoustic modes...''

\newpage

\vspace{7.cm}
Fig. 5 - T. Scopigno et al, ``High frequency acoustic modes...''


\begin{references}
\bibitem{hemmati} M. Hemmati, C.T. Moynihan, C.A. Angell, J. Chem.
Phys {\bf 115}, 6663 (2001).

\bibitem{nev-moy}  S. V. Nemilov, G. T. Petrovskii, and L. A.
Krylova, Inorg. Mater. (Transl of Neorg. Mater.) 4, 1453 (1968).
C. T. Moynihan and S. Cantor, J. Chem. Phys. 48, 115 (1968).

\bibitem {x} A. H. Narten, J. Chem. Phys. 56, 1905 (1972); A. J. Leadbetter
and A. C. Wright, J. Non-Cryst. Solids, 7, 156 (1972).

\bibitem{neu} A. C. Wright, A. G. Clare, G. Etherington, R. N. Sinclair,
S. A. Brawer, and M. J. Weber, J. Non-Cryst. Solids, 111, 139 (1989),
and references therein.

\bibitem{gal} F. L. Galeener, A. J. Leadbetter, and M. W. Stringfellow,
Phys. Rev. B 27, 1052 (1983).

\bibitem {kratt} J. T. Krause and C. R. Kurkjian, J. Am. Ceram. Soc. 51, 226
(1968).

\bibitem{note1} In the literature there is often confusion about the relation of
the sound absorption with the Brillouin linewidth. In particular, $\Gamma$ is
sometimes reported as the full width at half maximum while some other times is the
half width at half maximum. We believe that this apparent controversy is related
to the choice of the energy attenuation or of the field attenuation, respectively.
In this paper we always consider the energy attenuation.

\bibitem{noiV}  R. Verbeni, F. Sette, M. Krisch, U.Bergman, B. Gorges, C.
Halcoussis, K. Martel, C. Masciovecchio, J. F. Ribois, G. Ruocco and H. Sinn,
J. Of Synchrotron Radiation, {\bf 3}, 62 (1996).

\bibitem{noiM}  C. Masciovecchio, U. Bergman, M. Krisch, G. Ruocco, F. Sette
and R. Verbeni, Nucl. Inst. and Meth., {\bf B-111}, 181 and {\bf B-117}, 339
(1986).

\bibitem{bensil} M. Foret, E. Courtens, R. Vacher and J.B. Suck, Phys. Rev. Lett.
{\bf 77} 3831 (1996); P. Benassi, M. Krisch, C. Masciovecchio, V. Mazacurati,
G. Monaco, G. Ruocco, F. Sette and R. Verbeni, Phys. Rev. Lett. {\bf 77} 3835
(1996).

\bibitem{massil} C. Masciovecchio, V. Mazzacurati, G. Monaco, G. Ruocco, T. Scopigno,
F. Sette, P. Benassi, A. Cunsolo, A. Fontana, M. Krisch, A. Mermet, M.
Montagna, F. Rossi, M. Sampoli, G. Signorelli, R. Verbeni, Phil. Mag. B {\bf
79}, 2013 (1999).

\bibitem{matic} A. Matic, L. Borjesson, G. Ruocco, C. Masciovecchio, A. Mermet,
F. Sette and R. Verbeni, Europhys. Lett. {\bf 54}, 77 (2001).

\bibitem{baluc}  U. Balucani and M. Zoppi, Dynamics of the Liquid State
(Clarendon Press, Oxford, 1994).

\bibitem {kraspeed} This value was taken form G. E. Walrafen, Y. C. Chu, and M. S.
Hokmabadi, J. Chem. Phys. 92, 6987 (1990) where it was attributed to data by
J. T. Krause and C. R. Kurkjian.

\bibitem{har}  G. Ruocco, F. Sette, R. Di Leonardo, G. Monaco, M.
Sampoli, T. Scopigno and G. Viliani, Phys. Rev. Lett. {\bf84}, 5788 (2000).

\bibitem{litiosim}  T. Scopigno, G. Ruocco , F. Sette and G. Viliani,
to appear in Phys. Rev. E.

\bibitem {pilla} O. Pilla, A. Cunsolo, A. Fontana, C. Masciovecchio, G. Monaco,
M. Montagna, G. Ruocco, T. Scopigno, F. Sette, Phys. Rev. Lett.
{\bf85}, 2136 (2000).

\bibitem{zhu} T.C. Zhu, H.J. Maris and J. Tauc, Phys. Rev. B {\bf 44}, 4281 (1991).

\bibitem{vacher} R. Vacher, J. Pelous, E. Courtens, Phys. Rev. B {\bf 56}, R481
(1997).

\end{references}
\end{document}